\documentclass[conference]{IEEEtran}
\IEEEoverridecommandlockouts

\usepackage{tikz}
\usepackage{textcomp}
\usepackage{hyperref}
\usepackage{lipsum}
\usepackage{cite}
\usepackage{amsmath,amssymb,amsfonts}
\usepackage{algorithmic}
\usepackage{graphicx}
\usepackage{textcomp}
\usepackage{xcolor}
\usepackage{xspace}
\usepackage{balance}
\usepackage{multirow}
\def\BibTeX{{\rm B\kern-.05em{\sc i\kern-.025em b}\kern-.08em
    T\kern-.1667em\lower.7ex\hbox{E}\kern-.125emX}}

\newcommand\copyrighttext{%
  \footnotesize \textcopyright 2023 IEEE. Personal use of this material is permitted.
  Permission from IEEE must be obtained for all other uses, in any current or future 
  media, including reprinting/republishing this material for advertising or promotional 
  purposes, creating new collective works, for resale or redistribution to servers or 
  lists, or reuse of any copyrighted component of this work in other works. 
  DOI: \href{<http://tex.stackexchange.com>}{10.1109/TENCON58879.2023.10322536}}
\newcommand\copyrightnotice{%
\begin{tikzpicture}[remember picture,overlay]
\node[anchor=south,yshift=10pt] at (current page.south) {\fbox{\parbox{\dimexpr\textwidth-\fboxsep-\fboxrule\relax}{\copyrighttext}}};
\end{tikzpicture}%
}

\begin{document}
%
\copyrightnotice

\title{Design and Numerical Analysis of Hyperbolic Metamaterial Based Ultrasensitive \textit{E. Coli} Sensor\\}

\author{
\IEEEauthorblockN{Dip Sarker}
\IEEEauthorblockA{\textit{Department of Electrical and Electronic Engineering} \\
\textit{Bangladesh University of Engineering and Technology}\\
Dhaka, Bangladesh\\
0422062377@eee.buet.ac.bd}
\and
\IEEEauthorblockN{Ahmed Zubair*}
\IEEEauthorblockA{\textit{Department of Electrical and Electronic Engineering} \\
\textit{Bangladesh University of Engineering and Technology}\\
Dhaka, Bangladesh\\
*ahmedzubair@eee.buet.ac.bd}
}  
\maketitle

\begin{abstract}
We proposed an extremely sensitive \textit{E. Coli} sensor based on a hyperbolic metamaterial structure combining ultra-thin Ag-Al$_2$O$_3$ layers to minimize metallic optical loss. The principle relied on detecting the change in the resonance wavelength due to the interaction of bacteria with the surrounding aqueous environment by utilizing the finite-difference time-domain numerical technique. Our proposed hyperbolic metamaterial \textit{E. Coli} sensor operated in the range from visible to near-infrared wavelengths exhibiting strong bulk plasmon polaritons at the hyperbolic regime ($\lambda \geq$ 460 nm). An anisotropic hyperbolic range was obtained theoretically by solving the effective medium theory. An outstanding sensitivity of 9000 nm per bacteria was achieved for a bulk plasmon-polariton mode. The hyperbolic metamaterial was the origin of obtaining such extremely high sensitivity; no bulk plasmon polaritons were found without hyperbolic metamaterial. We analyzed the effect of different shapes in two-dimensional Ag differential grating on sensing performance. Additionally, we compared the performance parameters of our proposed \textit{E. Coli} sensor with recently demonstrated sensors. Our proposed hyperbolic metamaterial structure has the potential as a highly sensitive \textit{E. Coli} sensor operating in a wide range of wavelengths for label-free detection. 
\end{abstract}
\section{Introduction}
Even while freshwater is plentiful in populated nations like Bangladesh, it is frequently contaminated by different microbes. As the majority of the population relies on freshwater supplies, various water-born bacteria (i.e. \textit{Escherichia coli (E. Coli)}) in drinking water cause fatal diseases like diarrhea specifically in children in rural areas \cite{TOFAIL}. \textit{E. Coli} infections can result in severe stomach pains, diarrhea, and vomiting. In most cases, healthy adults recover from the infection within a week. Young children and older adults are especially vulnerable to a potentially fatal kind of renal failure. Thus, simple, easily-operable, real-time detection kits for \textit{E. Coli} are required in the point-of-care (POC) of rural areas. Sensors using the plasmonic phenomena have recently garnered the interest of academics due to their label-free and real-time detecting capabilities. However, smaller molecules and weights like \textit{E. Coli} are difficult to detect in very diluted solutions due to their lower polarizability\,\cite{Yan}. 

Holowko \textit{et al.} reported an \textit{E. Coli} biosensor to identify and eradicate \textit{Vibrio cholerae}, which can be utilized for cholera disease prevention and treatment. They performed their sensing principles based on a cholera autoinducer-1 (CAI-1) and the cqs-lux system \cite{Ho}. Lubkowicz \textit{et al.} proposed a biosensor to detect \textit{Staphylococcus aureus} by engineering \textit{Lactobacillus reuteri} with the help of agr system. Their proposed sensor senses autoinducer peptide-I which is produced by \textit{Staphylococcus sp.} during pathogenesis \cite{prem}. However, these biosensors require chemicals and expertise to handle which are expensive and sometimes inaccessible in POC\,\cite{ying}. Photonic crystal-based optical sensors can provide a better signal-to-noise ratio, sensitivity, and figure of merit (FOM) \cite{sharma}. However, these sensors need a complex fabrication procedure and show low anisotropic response during light-matter interaction\,\cite{Chowdhury}. On the contrary, metamaterials are a type of artificial material with unique electromagnetic characteristics that make them useful in diverse applications, including biosensing. Metamaterials-based sensors have unique properties to enhance sensitivity; however, microbes of small weight and size cannot be detected due to their lower polarizability. Additionally, these sensors provide a low FOM. Hyperbolic metamaterials (HMM) can resolve this problem and demonstrate exceptional sensitivity arising from enhanced light-matter interactions\,\cite{Gao}.

HMMs are artificial materials that exhibit hyperbolic dispersion, which means they have anisotropic optical properties. Specifically, they have a negative permittivity along one direction and a positive permittivity along another direction. This allows for unique optical properties such as strong anisotropy and the enhanced photonic density of states (theoretically infinity PDOS), enabling subwavelength imaging \cite{HMM book}. HMM-based sensors can detect small biomolecules with high sensitivity \cite{Sreekanth2016, Wang}. Moreover, these sensors are simple to construct, inexpensive, and have adjustable resonance properties compared to chemical and photonic sensors. There is a huge scope of research on designing HMM-based biosensors with enhanced performance by addressing the drawbacks of conventional biosensors. 
\begin{figure}[ht]
\centering
\includegraphics[width=9cm]{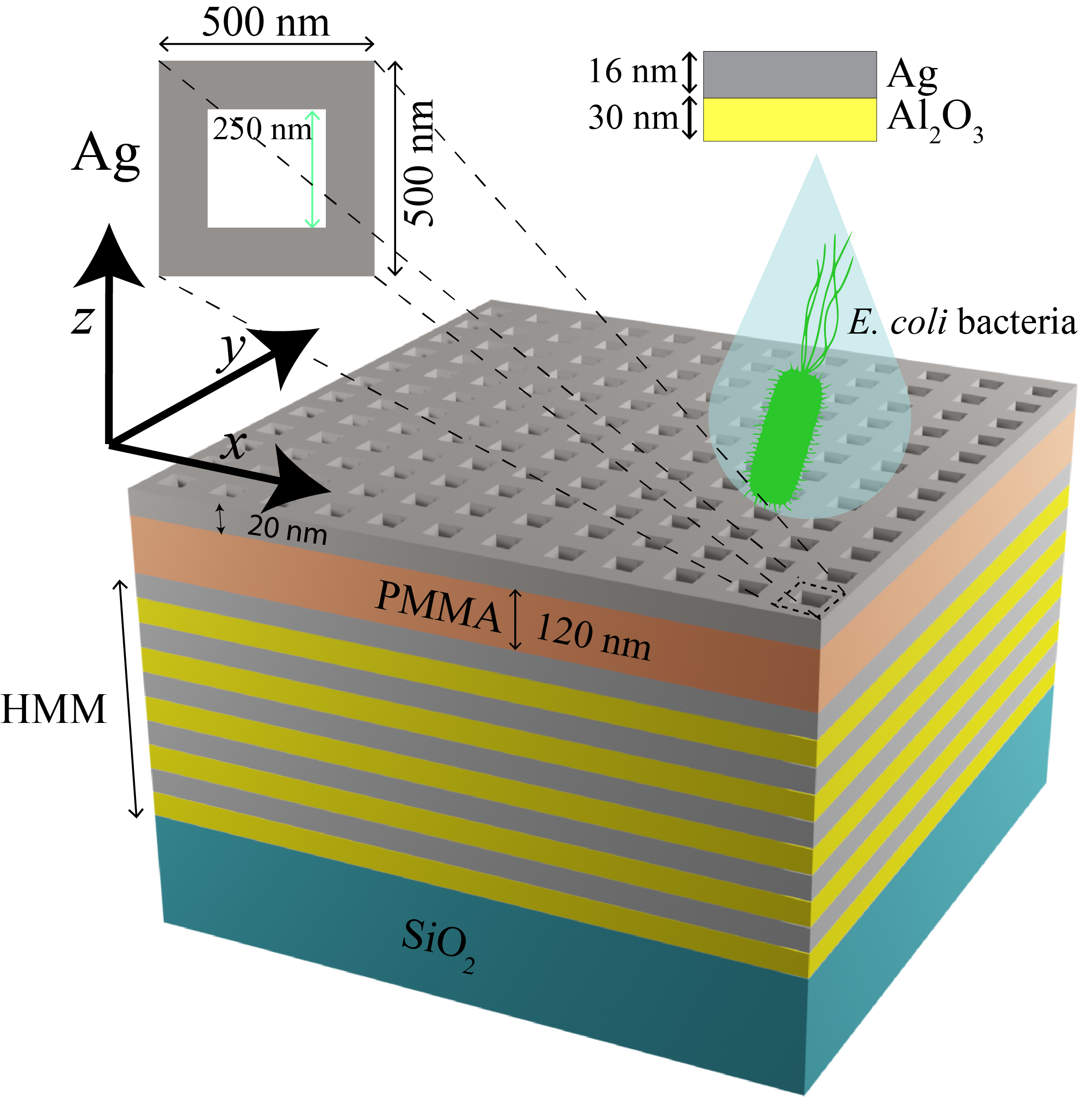}
\caption{Illustration of the proposed HMM-based periodic sensor structure. Inset shows a unit cell of rectangular 2D Ag diffraction grating hole with a period of 500 nm$\times$500 nm and rectangular air hole of 250 nm$\times$250 nm.}
\label{fig:1}
\end{figure}

In this work, we reported a portable \textit{E. Coli} sensor based on HMM comprising alternating Ag-Al$_2$O$_3$ layers coupled with periodic rectangular Ag diffraction gratings hole via a dielectric spacer. We investigated the reflectance of light from the proposed sensor using the finite-difference time-domain (FDTD) analysis technique. The hyperbolic region of the proposed HMM sensor was obtained by calculating effective medium theory (EMT). We observed the resonance wavelength shift for bio-sensing applications due to the change in refractive index (RI) and calculated the sensitivity for the proposed HMM sensor. We comprehensively studied the optical properties of the proposed HMM sensor structure. The effects of the incidence angle of light on the proposed structure and different shapes in the diffraction grating layer were analyzed. Moreover, a comparative performance analysis of our sensor and previously demonstrated biosensors were conducted.  
\section{Device design \& simulation methodology}
We proposed an HMM structure comprising alternating Ag-Al$_2$O$_3$ layers. A periodic rectangular two-dimensional (2D) Ag diffraction gratings hole coupled with HMM by a dielectric spacer of polymethyl methacrylate (PMMA), as illustrated in Fig.~\ref{fig:1}. Inset shows the schematic illustration of a unit cell of the rectangular Ag diffraction grating hole.  Rectangular 2D grating hole structure with a period of 500 nm and a hole size of 250 nm$\times$250 nm was used to achieve highly sensitive sensors in the range of visible to near-infrared regions. The thickness of Ag diffraction gratings was set to 20 nm. The HMM had 16 alternating layers of Ag-Al$_2$O$_3$ with thicknesses of Ag and Al$_2$O$_3$ in each layer were 16 nm and 30 nm, respectively. We modeled the total relative permittivity, $\epsilon_{total}$ of Ag by employing the Drude model \cite{Drude},
\begin{equation}
    \epsilon_{total} = \epsilon - \frac{\omega_p}{\omega^2 + i \omega \Gamma}
    \label{Eq. 1}
\end{equation}
Here, $\epsilon$, $\omega_p$, $\omega$, and $\Gamma$ are the permittivity, plasma resonance frequency, angular frequency, and plasma collision, respectively. We adopted $\epsilon$ of 1, $\omega_p$ of 9 eV, and $\Gamma$ of 0.07 eV in our study \cite{Kravets}. The refractive indices, $n$ of Al$_2$O$_3$ was modeled by using Sellmeier equation\,\cite{Malitson},
\begin{equation}
    n^2(\lambda) = 1+ \frac{B_1 \lambda^2}{\lambda^2 - C_1} + \frac{B_2 \lambda^2}{\lambda^2 - C_2} + \frac{B_3 \lambda^2}{\lambda^2 - C_3}
    \label{Eq. 2}
\end{equation}
Here, $\lambda$ is the wavelength of light. $B_1$, $B_2$, $B_3$, $C_1$, $C_2$, and $C_3$ are the Sellmeir coefficients. We utilized $B_1$ of 1.4313493, $C_1$ of 0.0726631 $\mu$m$^2$, $B_2$ of 0.65054713, $C_2$ of 0.1193242 $\mu$m$^2$, $B_3$ of 5.3414021, and $C_3$ of 18.028251 $\mu$m$^2$ in our work\,\cite{Malitson}. The optical properties of SiO$_2$ were taken from Palik\,\cite{Palik}. PMMA's refractive index of 1.49\,\cite{Beadie} with a thickness of 120 nm was used as a dielectric spacer between HMM and rectangular 2D Ag diffraction grating hole. The optical properties of single \textit{E. Coli} bacteria (difference of refractive index, $\Delta$ = 0.005 RIU) were collected from the optofluidic immersion refractometry technique during the performance analysis of the sensor\,\cite{Liu}.

\begin{figure}[ht]
\centering
\includegraphics[trim={0.32cm 0cm 0cm 1.2cm}, clip, width=9cm]{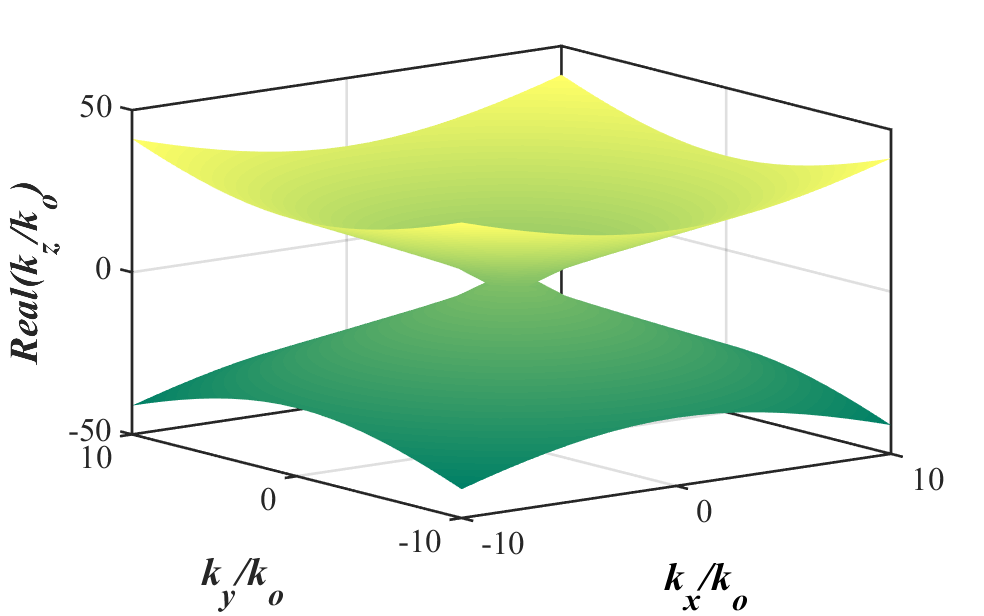}
\caption{The calculated iso-frequency surface curve for our proposed multilayered hyperbolic metamaterial. Here, $\epsilon_{\parallel} > 0$ and $\epsilon_{\perp} < 0$ are adopted.}
\label{dispersion}
\end{figure}

We monitored the reflectance of light from the proposed structure using a three-dimensional (3D) FDTD analysis technique. Periodic boundary conditions were adopted in the x- and y-directions, and 12 steep-angle perfectly matched layers (PMLs) were utilized to absorb light out from the simulation region. A transverse magnetic (TM) polarized plane wave with an incidence angle of 30$^{\circ}$ was incident on the proposed structure along the negative z-direction. TM polarized light on our proposed structure provided better performance than transverse electric (TE) polarized light, as illustrated in Fig.~\ref{No HMM}(a). Therefore, we considered TM polarized light for our study. 3D FDTD simulation is a time-consuming and high memory requirements technique. Therefore, the non-uniform mesh was employed to conduct our numerical studies within limited memory storage and time. Simulations were stopped when the energy of the simulation appeared at 10$^{-5}$ of its initial energy. The temperature was set to be 300 K in our study. 
   
\begin{figure}[ht]
\centering
\includegraphics[width=8.5cm]{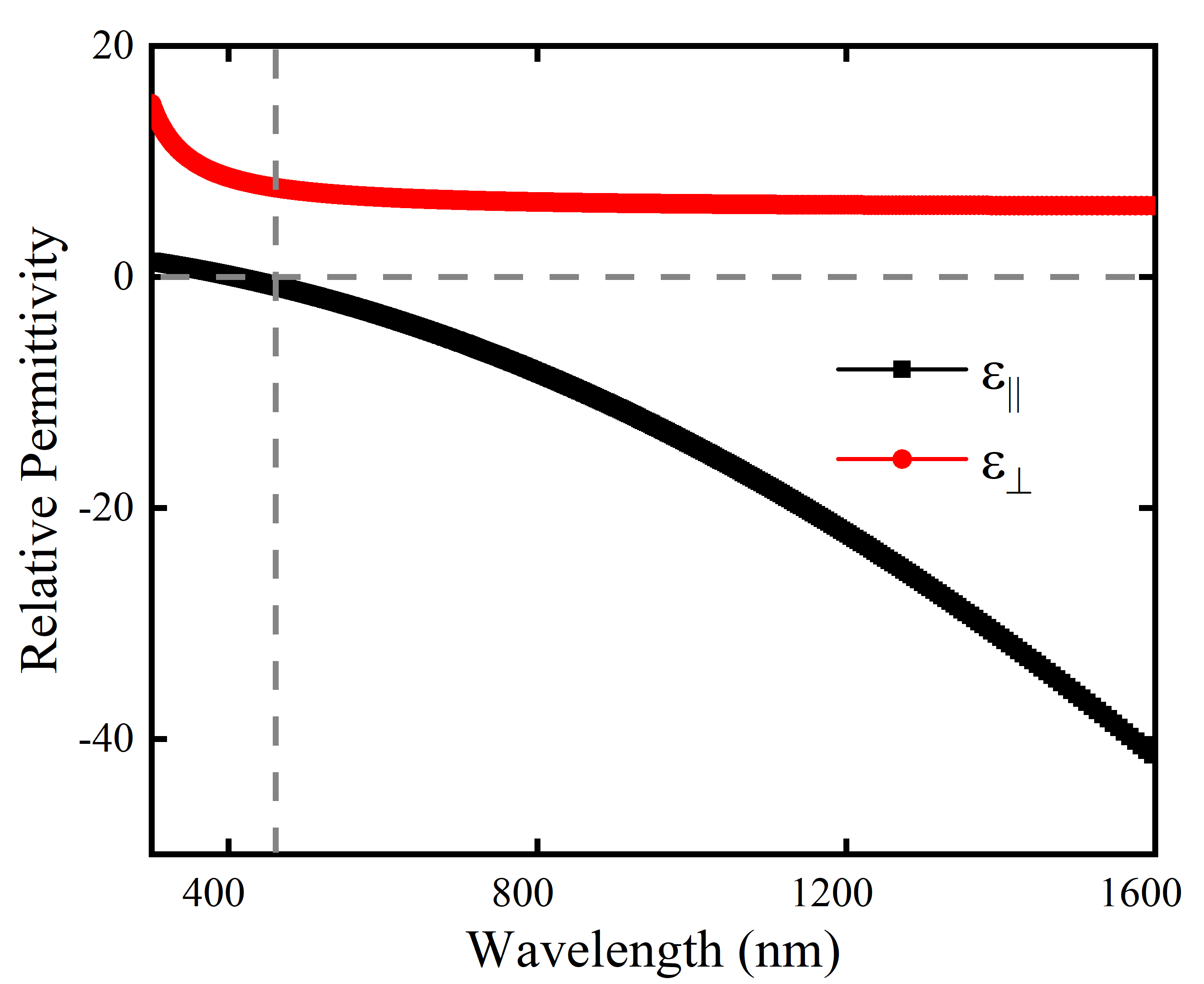}
\caption{The calculated relative permittivity of Ag--Al$_2$O$_3$ HMM structure using effective medium theory. Hyperbolic dispersion occurs when $\epsilon_{\parallel}<0$ and $\epsilon_{\bot}>0$ ($\lambda \geq$ 460 nm).}
\label{fig:2}
\end{figure}
\section{Results \& Discussions}
\subsection{Properties of HMM}
Most materials are isotropic and have spherical iso-frequency contours. However, when materials exhibit anisotropy or do not react uniformly in all directions, this spherical contour turns into an ellipsoid. The dispersion relation of such materials can be described as\,\cite{HMM book},
\begin{equation}
    \frac{k_x^2}{\epsilon_{\perp}} + \frac{k_y^2}{\epsilon_{\perp}} + \frac{k_z^2}{\epsilon_{\parallel}} = k_o^2
    \label{Eq. hp1}
\end{equation}
Here,  $k_x$, $k_y$, and $k_z$ represent the wave vector in x-, y-, and z-directions, respectively. $\epsilon_{\perp}$ and $\epsilon_{\parallel}$ are the parallel and perpendicular components of permittivity. $k_o = \omega/c$ is the wave vector in free space. HMM is an anisotropic medium where properties are different in all directions. Type-1 and type-2 are the two classifications of HMM, depending on their dielectric tensor. Type-1 HMM provides a hyperbolic region at $\epsilon_{\parallel}>0$ and $\epsilon_{\bot}<0$ whereas a hyperbolic region is found at $\epsilon_{\parallel}<0$ and $\epsilon_{\bot}>0$ for type-2 HMM. Fig.~\ref{dispersion} represented the isofrequency surface plot of our proposed type-2 HMM. The isofrequency surface was calculated using the Eq.~\ref{dispersion} of the dispersion relation. Our proposed HMM exhibited hyperbolic dispersion, with permittivity having a negative sign. A strong hyperbolic dispersion was observed at $\lambda \geq$ 460 nm where $\epsilon_{\parallel}<0$ and $\epsilon_{\bot}>0$ that sustain highly confined plasmon-guided modes across a wide visible-to-near-infrared wavelength range, as can be seen in Fig.~\ref{fig:2}. The parallel and perpendicular permittivity components of our proposed HMM structure were calculated by using EMT\,\cite{jacob}, 
\begin{equation}
    \epsilon_{\parallel} = \frac{t_m \epsilon_m + t_d \epsilon_d}{t_m + t_d}
    \label{Eq. 3}
\end{equation}
\begin{equation}
    \epsilon_{\perp} = \frac{\epsilon_m \epsilon_d (t_m + t_d)}{t_m \epsilon_d + t_d \epsilon_m}
    \label{Eq. 4}
\end{equation}
Here, $\epsilon_m$ and $\epsilon_d$ are the permittivities of Ag and Al$_2$O$_3$, respectively. And, $t_m$ and $t_d$ are the thicknesses of Ag and Al$_2$O$_3$, respectively. 
\subsection{Dependency of HMM}
\begin{figure}[ht]
\centering
\includegraphics[trim={0.1cm 0cm 0.2cm 0cm}, clip, width=8.5cm]{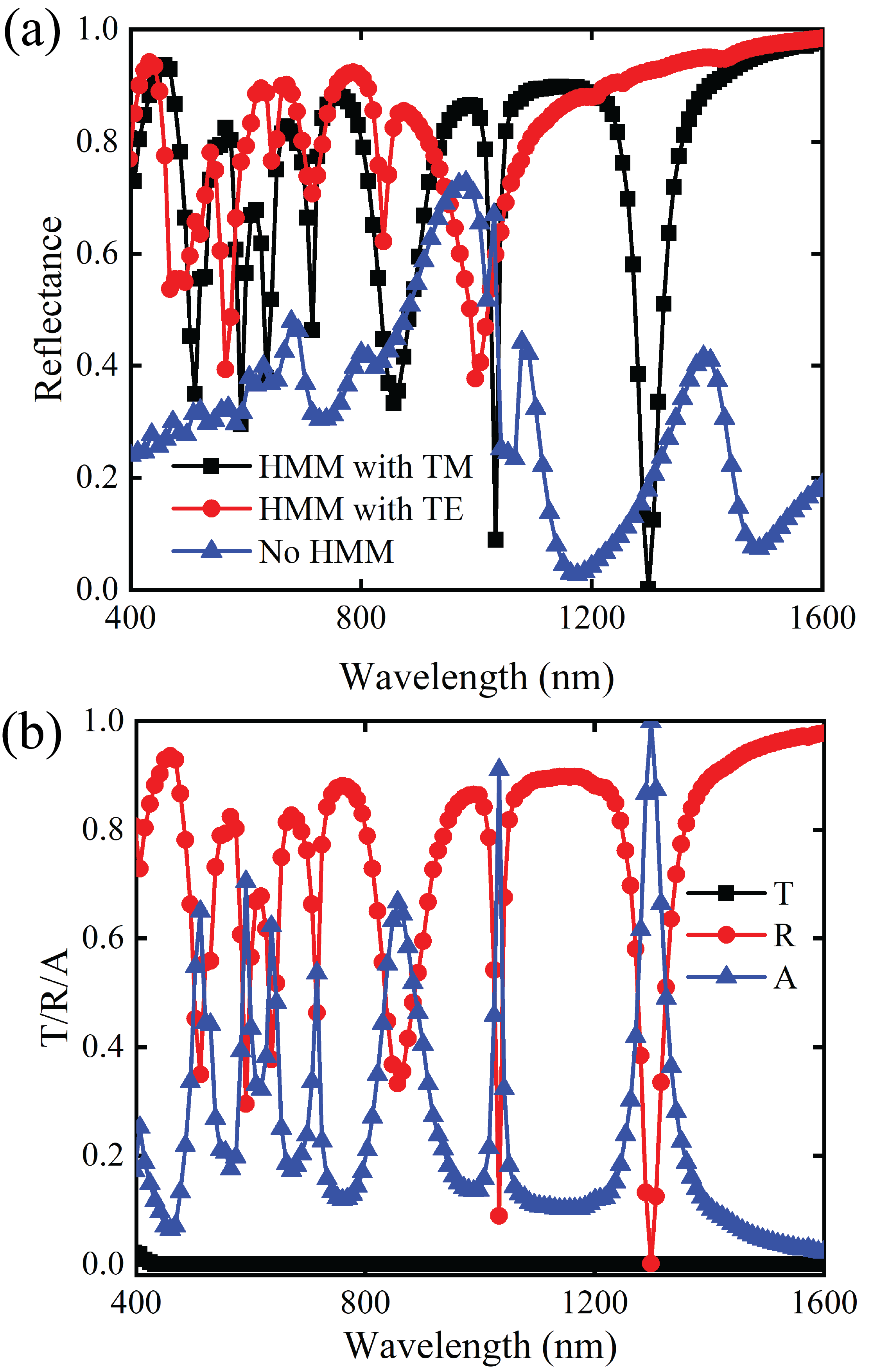}
\caption{(a) Simulated reflectance spectra of the sensor with and without HMM. HMM with TM polarized light exhibits strong BPP modes at resonant frequencies. No bulk plasmon polariton modes are observed without HMM. HMM provides BPP modes with high-quality factors. (b) Transmittance (T), reflectance (R), and absorptance (A) spectra of our proposed HMM \textit{E. Coli} sensor.}
\label{No HMM}
\end{figure}

When TM polarized plane wave was incident on the proposed structure, bulk plasmon-polariton (BPP) modes were obtained at resonant wavelengths of 512 (BPP 1), 591 (BPP 2), 635 (BPP 3), 715 (BPP 4), 856 (BPP 5), 1033 (BPP 6), and 1298 (BPP 7) nm in the range of visible to near-infrared wavelength, as shown in Fig.~\ref{No HMM}(a). Additionally, we varied the HMM layer number and determined the impact on the reflectance spectra of our proposed sensor structure. For fewer HMM layers, we obtained unstable BPP modes, which alter with increasing the number of HMM layers. The alternating 16 Ag and Al$_2$O$_3$ layers provided stable BPP modes with high-quality factors, Q. The Q is given by,
\begin{equation}
    Q = \frac{\lambda_o}{\mbox{FWHM}}
    \label{Q}
\end{equation}
Here, $\lambda_o$ and FWHM denote the resonance wavelength and full-width half maximum, respectively. High-Q factors of 107.6 and 32.45 were enumerated at BPPs 6 and 7, respectively, which were significantly higher compared to those of previous reports\,\cite{Chowdhury, Ahmed}. No BPP modes can be achieved without the HMM layers (see Fig.\,\ref{No HMM}(a)). It can be inferred that the origin of the BPP modes was HMM which supported high-quality factors. Incoming light was absorbed in resonant frequencies due to light-matter interactions, as depicted in Fig.~\ref{No HMM}(b). Transmission through our proposed HMM biosensor was negligible.  
\begin{figure}[ht]
\centering
\includegraphics[width=8cm]{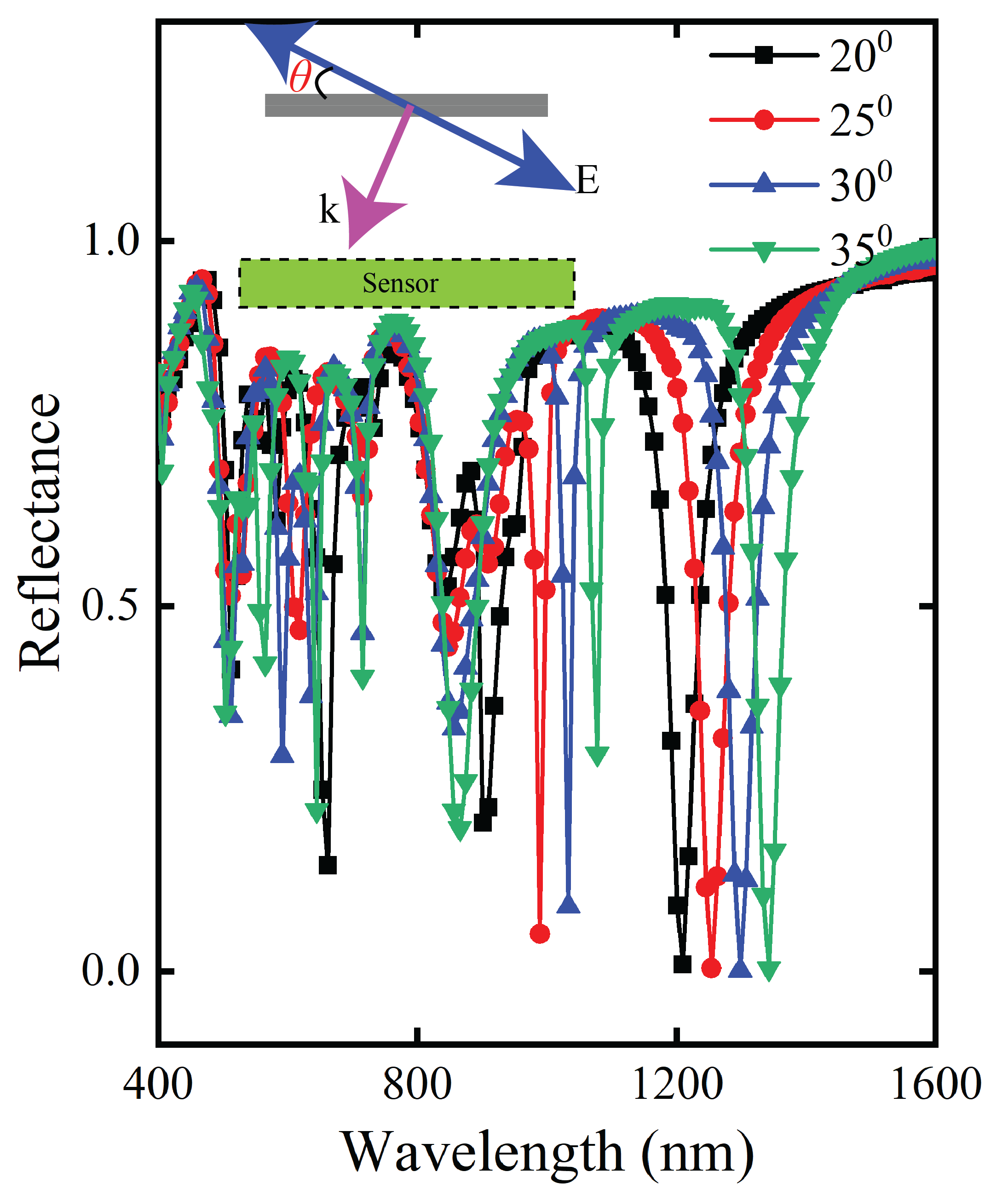}
\caption{Reflectance spectra of the proposed sensor structure at different light incidence angles. We observed the anisotropy behavior of HMM by the shift in resonance wavelength. Inset shows the incident angle ($\theta$) of the TM polarized plane wave.}
\label{angles}
\end{figure}
\subsection{Effect of incidence angle}
The structure exhibited anisotropic behavior ($\lambda \geq$ 460 nm) due to the change in the effective index of plasmonic guided modes at different incidence angles of TM light, as illustrated by red-shift resonance in Fig.~\ref{angles}. Inset depicted the incidence angle of the plane wave where $\theta$, E, and k represent the incidence angle, electric field, and propagation constant of TM polarized light. Here, the RI of water was utilized for the surrounding medium of all simulations. Seven reflectance dips were found in the hyperbolic region ($\geq$ 460 nm), which represented high-k modes. The high-k modes had wavelengths of 512, 591, 635, 715, 856, 1033, and 1298 nm for the incidence angle of 30$^{\circ}$. As there was zero transmission through the suggested HMM structure, BPP 6 and BPP 7 exhibited 91\% and 99.9\% absorption, whereas other BPP modes exhibited roughly 60\% absorption.   

\subsection{Impact of shapes in diffraction grating}
\begin{figure}[ht]
\centering
\includegraphics[width=8.5cm]{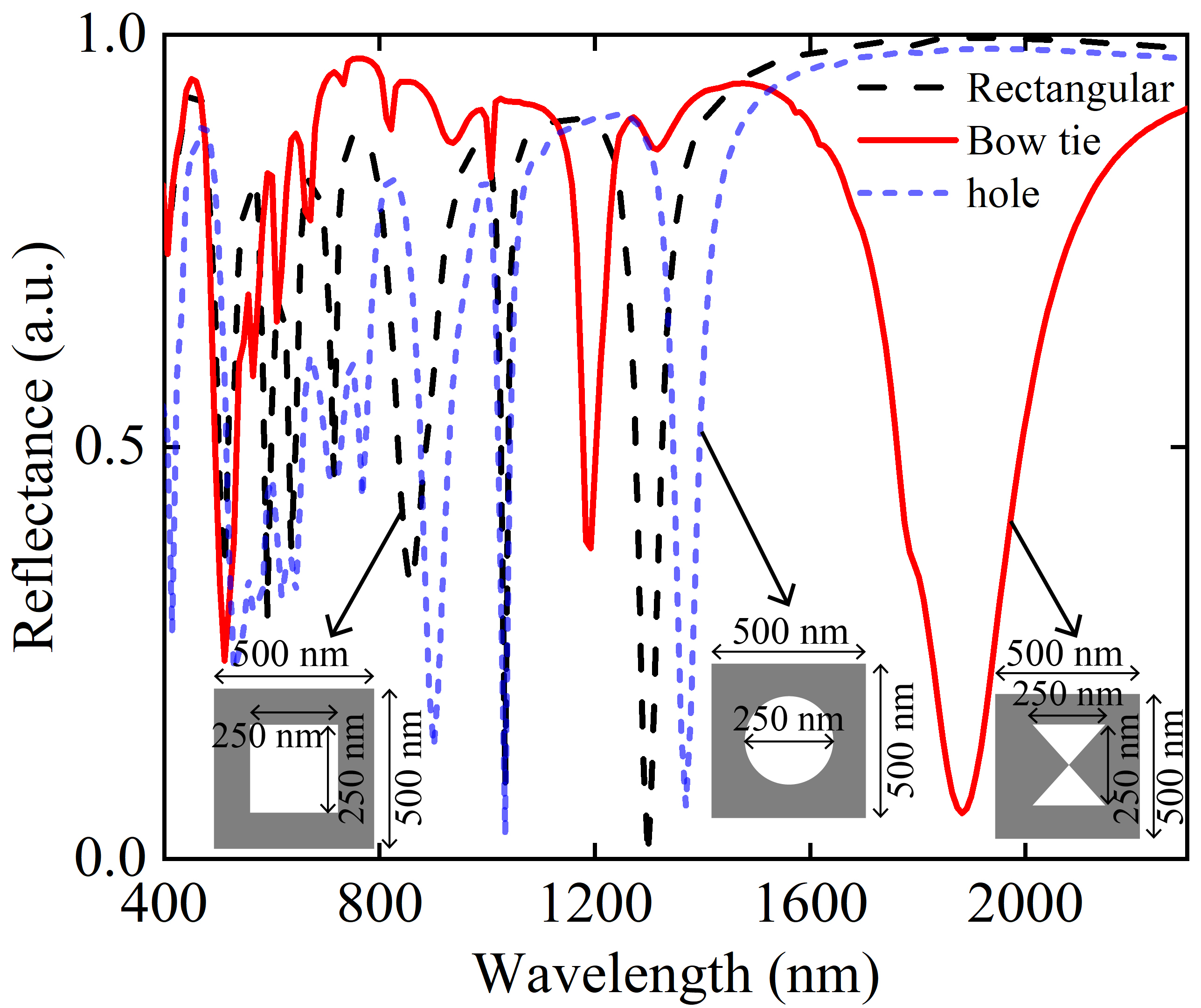}
\caption{Reflectance spectra for various shapes of Ag diffraction arrays. Rectangular Ag diffraction grating provides better performance than other shapes of Ag diffraction grating. Insets represent the schematic illustration of unit cells with rectangular, circular, and bow tie hole Ag diffraction grating.}
\label{fig:4}
\end{figure}

We analyzed the effect of different shapes of the 2D Ag diffraction grating layer. Bow tie and hole shapes were considered on Ag diffraction grating with parameter optimization. The circle's radius in hole shape was set to be 125 nm, whereas the other parameters were similar to rectangular Ag diffraction grating. The bow tie shape had two inverted air triangles with a length of 250 nm in the x-direction and a width of 250 nm in the y-direction, and other parameters remained the same. We obtained weak plasmon polariton coupling with light in the visible wavelength range for the circular hole Ag grating. We found BPP modes beyond the visible range, as depicted in Fig.~\ref{fig:4}. Weak BPP modes were obtained at lower wavelengths around 530 nm, and three BPP modes with narrow FWHM were achieved at resonant wavelengths of 900, 1033, and 1369 nm. Though bow tie Ag 2D grating provided wide coverage in the visible to infrared wavelength range, BPP modes reduced the amplitude of the signal at the resonant frequencies of BPP modes. Three low-Q BPP modes were obtained at 512, 1292, and 1881 nm resonance wavelengths. On the contrary, a rectangular Ag 2D grating structure can provide not only a narrow full-width half maximum but also a high number of BPP modes covering from visible to infrared wavelength range. Thus, we employed a rectangular Ag 2D grating structure to design our highly sensitive biosensor.  
\subsection{Sensing performance}
\begin{figure}[ht]
\centering
\includegraphics[width=8.8cm]{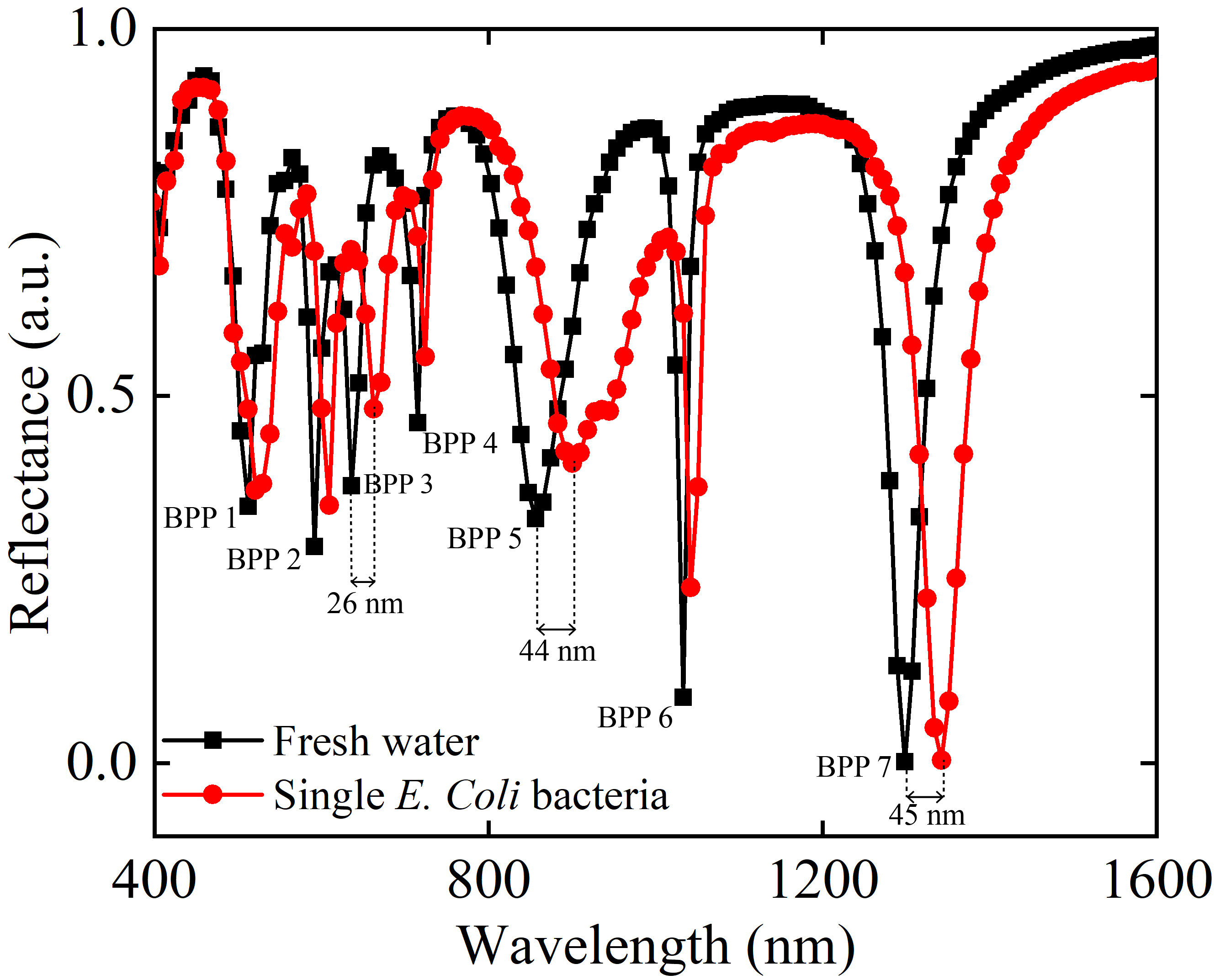}
\caption{Red-shifted reflectance spectra of the HMM sensor by injecting single \textit{E. Coli} bacteria containing fresh water.}
\label{fig:5}
\end{figure}

\begin{figure}[ht]
\centering
\includegraphics[trim={1.27cm 0cm 1.5cm 1cm}, clip, width=9.5cm]{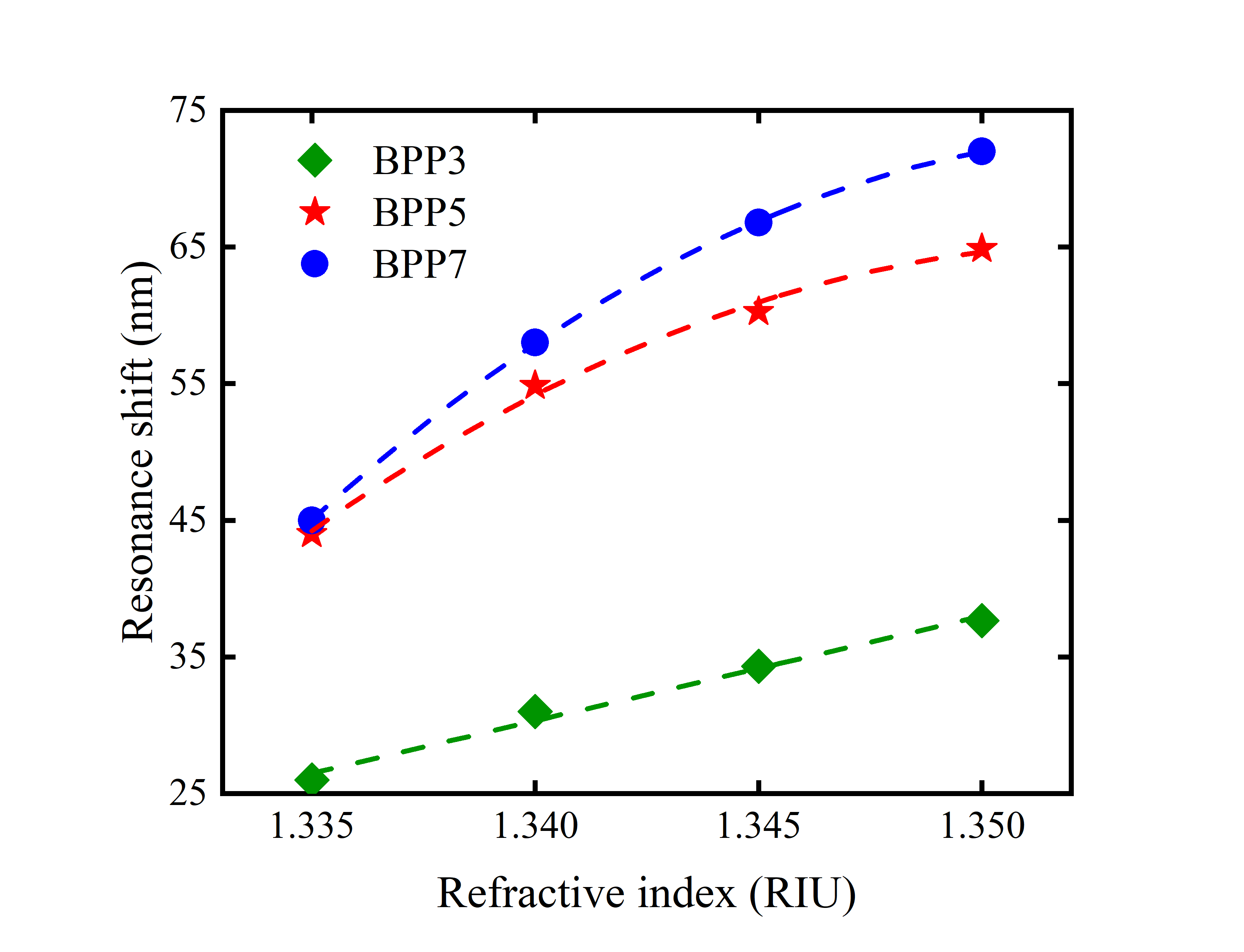}
\caption{Resonance behavior of our proposed biosensor in the presence of multiple \textit{E. Coli} molecules in aqueous solutions. The dash lines represent the fitted shift of BPP3, BPP5, and BPP7.}
\label{RI change}
\end{figure}

To determine the sensing capability of our proposed sensor, a small change in RI due to injecting single \textit{E. Coli} bacteria into aqueous fresh water solutions resulted in the shift of the reflectance spectrum. In addition, a comparison between the reflected light of water and single \textit{E. Coli} bacteria medium was studied to observe the wavelength shift and calculated the sensitivity of our proposed sensor structure. The optical properties of \textit{E. Coli} bacteria were collected from the optofluidic immersion refractometry technique of Liu \textit{et. al.} \cite{Liu}. We obtained seven BPP modes at resonance frequencies and calculated the shift of reflectance spectra for single \textit{E. Coli} bacteria in hyperbolic region ($\lambda \geq$ 460 nm), as shown in Fig.~\ref{fig:5}. The sensitivity, S was enumerated by\,\cite{Ye},
\begin{equation}
    S = \frac{\Delta \lambda}{\Delta n}
    \label{Eq. 5}
\end{equation}
Here, $\Delta \lambda$ and $\Delta n$ are the shift of resonance wavelength and the change of bacteria concentration, respectively. The resonance wavelengths of BPP 3, 5, and 7 were shifted by 26, 44, and 45 nm, respectively, and can be considered bacteria-sensitive modes. The calculated sensitivities for these three modes were 5200 nm per bacteria, 8800 nm per bacteria, and 9000 nm per bacteria, respectively. 
The effect of the presence of multiple \textit{E. Coli} bacteria were studied in our proposed sensor, as shown in Fig.~\ref{RI change}. We conducted additional studies for the presence of 2, 3, and 4 \textit{E. Coli} bacteria which corresponded to various values of RI. Various RI values reflected the concentration of \textit{E. Coli} bacteria existence in solution. RI values for 2,3, and 4 \textit{E. Coli} bacteria in water solution were 1.34, 1.345, and 1.35, respectively. We obtained shift in resonance wavelength with increasing the amount of \textit{E. Coli} bacteria in freshwater solution. A maximum resonance shift of $\sim$72 nm was observed for four \textit{E. Coli} bacteria in water solution. Based on our study, we formulated an empirical equation of BPP 7 for the RI dependence of resonance wavelength shift, $\delta \lambda_{\circ}$ given by,
\begin{equation}
    \delta \lambda_{\circ} = -142567.1 + 210688.6 n_{bac} - 77800 n_{bac}^2
    \label{empirical}
\end{equation}
Here, $n_{bac}$ denotes the RI of \textit{E. Coli} bacteria. The adjusted R$^2$ for the empirical equations was 0.9998.

\begin{figure}[ht]
\centering
\includegraphics[trim={1.75cm 1.25cm 1.25cm 1.9cm}, clip, width=9.5cm]{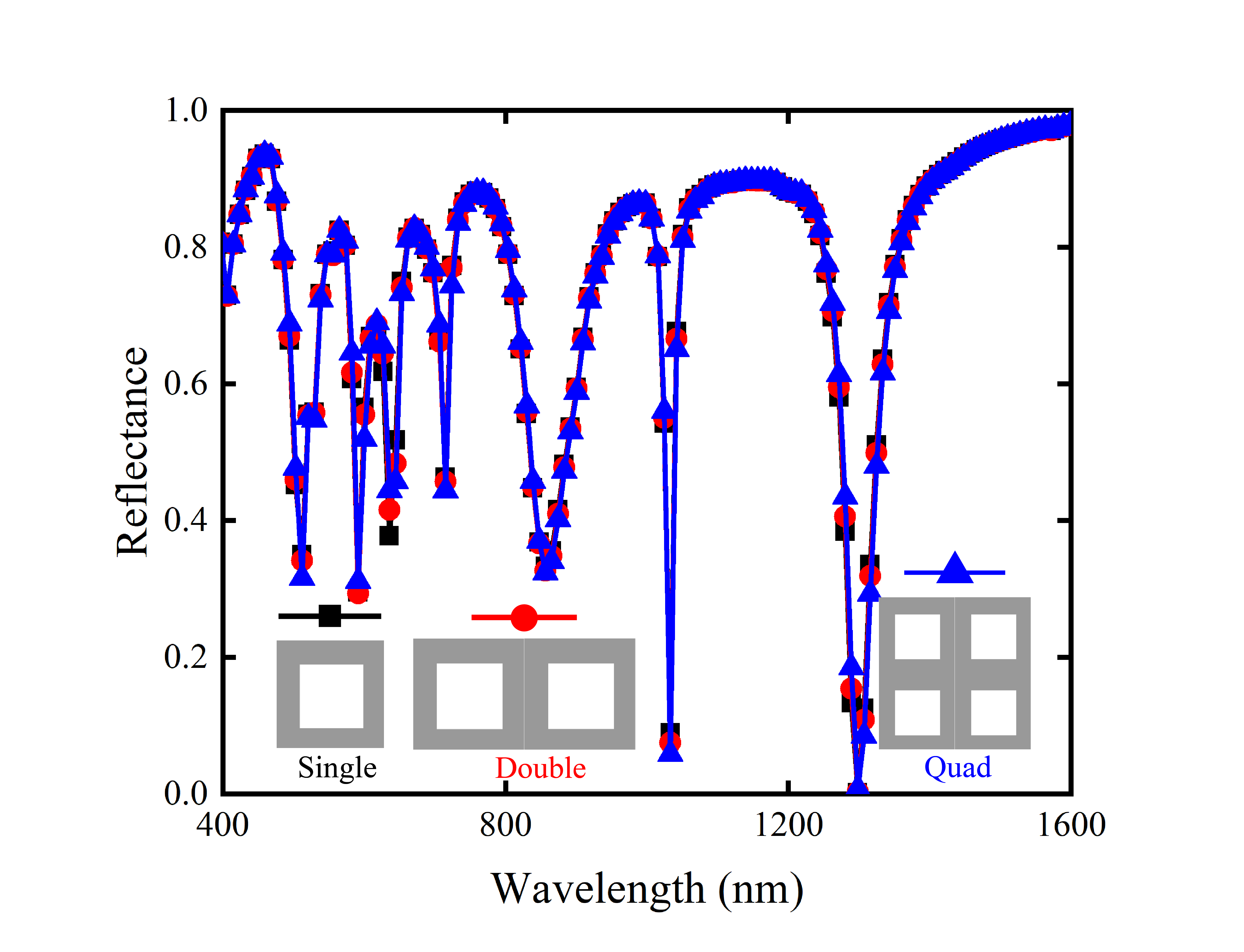}
\caption{Reflection spectra for our proposed single, double, and quad cell where the refractive index is 1.33. Insets depict the illustration of the unit cell.}
\label{cell number}
\end{figure}

A single microbe does not necessarily surround a unit cell of the sensor structure; rather, it may cover two or four cells simultaneously. Therefore, we studied multiple cells' impact on the HMM biosensor's reflection spectra. Here all the structural parameters remained the same as before, and the refractive index of the surroundings was set to 1.33 RIU. The amplitude of the reflection changed in some BPP modes; however, the resonant frequencies of BPP modes were the same for all three cases, as shown in Fig.~\ref{cell number}. A detailed comparative analysis of performance parameters among previously reported works and our work is enlisted in Table.~\ref{Tab:1}. Our proposed HMM structure exhibited extraordinary performance compared to existing demonstrated works. 
\begin{table*}[htbp]
\caption{Comparative performance analysis of biosensors}
\begin{center}

\begin{tabular}{c c c c c}
\hline
Structure & Wavelength ($\mu$m) & Sensitivity &  Q-factor & Ref. \\
\hline
Triangular gold nanoparticles & 0.5 -- 0.7 & $<$110.7 nm/mM & $<$ 2.92 & \cite{Chowdhury} \\

Au/1D photonic crystal & 0.6 -- 0.8 & 53.5 nm/RIU & $<$ 0.69 & \cite{Ahmed} \\

Double spilt-ring resonator & 150 -- 300 & 387 GHz/RIU & -- & \cite{Nine} \\ 

Graphene/MoS2 crumpled sensor & 4 -- 10 & 7499 nm/RIU & 67 & \cite{vahid} \\

Ag-Al$_2$O$_3$ HMM sensor & 0.5 -- 2.0 & 9000 nm/bacteria & 107.6 & This work \\
\hline
\end{tabular}
\label{Tab:1}
\end{center}
\end{table*}
\section{Conclusions}
We designed and numerically studied HMM-based highly sensitive biosensors in visible and infrared wavelength ranges in this paper. Using EMT, a theoretically calculated strong hyperbolic region at $\lambda \geq$ 460 nm was obtained. Shifting resonance dips by varying incidence angles proved the anisotropic property of HMM. We demonstrated a comparative analysis of reflectance spectra among different Ag grating shape structures and selected the sensor structure that provides narrow FWHM and highly sensitive response. Varying refractive index by injecting \textit{E. Coli} bacteria into aqueous solutions caused the shift of the reflectance spectrum. A maximum sensitivity of 9000 nm per bacteria was obtained for our proposed HMM sensor. The impact of multiple bacteria in water solutions and multiple cells in the simulation region was performed. Therefore, minuscule living organisms like water-borne bacteria can be detected using our proposed HMM sensor with high sensitivity.
\section*{Acknowledgments}
We thank the Robert Noyce Simulation Laboratory at the Department of Electrical and Electronic Engineering, Bangladesh University of Engineering and Technology (BUET) for providing technical support during the study. Dip Sarker acknowledges financial support from BUET through its Postgraduate Fellowship Program. Ahmed Zubair acknowledges the Basic Research Grant (Sonstha/R-60/Ref-4747) the Bangladesh University of Engineering and Technology provided. We acknowledge Partha Pratim Nakti for assisting in the illustration work.

\balance

\vspace{12pt}

\end{document}